\documentclass[runningheads]{llncs}
\usepackage[T1]{fontenc}
\usepackage{graphicx}
\usepackage{package}
\usepackage{cite}
\usepackage{amsmath,amssymb,amsfonts}
\usepackage{algorithmic}
\usepackage{graphicx}
\usepackage{textcomp}
\usepackage{xcolor}
\usepackage{multicol}

\begin{document}

\title{REST in Pieces: RESTful Design Rule Violations in Student-Built Web Apps
\thanks{This work has been partially supported by the Italian PNRR
MUR project PE0000013-FAIR.}
} 

\titlerunning{REST in Pieces}
%
\author{
Sergio Di Meglio \inst{1}
\and
Valeria Pontillo \inst{2,3}
\and
Luigi Libero Lucio Starace \inst{1}
}
\authorrunning{S. Di Meglio et al.}
%
\institute{
Department of Electrical Engineering and Information Technology, University of Naples Federico II, Italy\\
\email{sergio.dimeglio@unina.it, luigiliberolucio.starace@unina.it}\\
\and
Software Languages (SOFT) Lab, Vrije Universiteit Brussel, Belgium
\and
Gran Sasso Science Institute (GSSI), L'Aquila, Italy\\
\email{valeria.pontillo@gssi.it}
}
\maketitle              
\begin{abstract}

In Computer Science Bachelor's programs, software quality is often underemphasized due to limited time and a focus on foundational skills, leaving many students unprepared for industry expectations. To better understand the typical quality of student code and inform both education and hiring practices, we analyze 40 full-stack web applications developed in a third-year Web Technologies course. 
Using an automated static analysis pipeline, we assess adherence to REST API design rules. Results reveal frequent violations of foundational conventions, such as missing hyphens in endpoint paths (98\%), incorrect pluralization (88\%), and misuse of HTTP methods (83\%). These findings highlight the need for more focused instruction on API design and support the adoption of automated tools to improve code quality in student projects.

\end{abstract}

\section{Introduction}

In Computer Science Bachelor's programs, software development education often emphasizes functional application building over software quality, consistent with ACM guidelines recommending limited early focus on quality aspects \cite{clear2019computing}. Time constraints and difficult pedagogical trade-offs further limit the comprehensive teaching of software quality principles \cite{de2024automatic}. Moreover, empirical data on the structural and architectural quality of realistic student projects—especially full-stack applications—are scarce, hindering informed curriculum decisions and impacting industry readiness \cite{wlodarski2021assessment}.

This gap has significant industry implications: graduates lacking solid skills in clean, maintainable code shift the retraining burden to employers, who must invest resources to bring junior developers up to production standards \cite{10803360}. Bridging this academia-industry divide requires a better understanding of student software quality.

To address this gap, we conducted an empirical investigation on 40 full-stack web applications developed by students enrolled in a third-year Web Technologies course within a Computer Science Bachelor's program. This course, typically taken in the final semester before graduation, requires students to design and implement complete web systems that integrate both front-end and back-end components. To this end, we designed a semi-automated analysis pipeline that leverages \textsc{RESTRuler}, a tool to identify REST API design rule violations, an indicator of structural degradation and design decay, commonly associated with the accumulation of social debt in collaborative software projects.
Our findings indicate that while students delivered functional full-stack applications, they frequently violated key REST design conventions—most notably the use of hyphens in endpoint paths (98\% of projects), plural nouns for collections (88\%), and correct use of GET methods (83\%). 

These insights can inform teaching strategies to better prepare students for professional development. A replication package with all analysis artifacts is available to support further research and validation \cite{replicationpackage}.


\section{Background and Related Works} \label{sec:bk-related}

In the following, we introduce key concepts related to software quality, with a focus on design-level aspects specific to RESTful APIs. We also review existing tools for assessing REST API design rule violations and summarize prior research on the quality of student-developed software systems.

\begin{table}[!ht]
    \centering
    \small
    \setlength{\aboverulesep}{0pt}
    \setlength{\belowrulesep}{0pt}
    \setlength{\extrarowheight}{.5ex}
    \arrayrulecolor{black}
  \caption{The 14 rules implemented to \textsc{RESTRuler} \cite{RestRuler}.}
    \begin{tabularx}{\textwidth}{|X|l|l|}
    \toprule
    \rowcolor{arsenic}
    \textcolor{white}{\textbf{Rule Description}} & \textcolor{white}{\textbf{Identifier}} & \textcolor{white}{\textbf{Category}} \\
    \rowcolor{white}
    \texttt{401 ("Unauthorized")} must be used when there is a problem with the client's credential & RC401 & HTTP Status Codes \\
    \rowcolor{gray!5}
    A plural noun should be used for collection or store names & PluralNoun & URI Design \\
    \rowcolor{white}
    A singular noun should be used for document names & SingularNoun & URI Design \\
    \rowcolor{gray!5}
    A trailing forward slash (/) should not be included in URIs & NoTrailingSlash & URI Design \\
    \rowcolor{white}
    A verb or verb phrase should be used for controller names & VerbController & URI Design \\
    \rowcolor{gray!5}
    CRUD function names should not be used in URIs & NoCRUDNames & URI Design \\
    \rowcolor{white}
    Content-Type must be used & ContentType & Metadata Design \\
    \rowcolor{gray!5}
    Description of request should match with the type of the request & DescriptionType & Metadata Design \\
    \rowcolor{white}
    Forward slash separator (/) must be used to indicate a hierarchical relationship & ForwardSlash & URI Design \\
    \rowcolor{gray!5}
    \texttt{GET} and \texttt{POST} must not be used to tunnel other request methods & NoTunnel & Request Methods \\
    \rowcolor{white}
    \texttt{GET} must be used to retrieve a representation of a resource & GETRetrieve & Request Methods \\
    \rowcolor{gray!5}
    Hyphens (-) should be used to improve the readability of URIs & Hyphens & URI Design \\
    \rowcolor{white}
    Lowercase letters should be preferred in URI paths & Lowercase & URI Design \\
    \rowcolor{gray!5}
    Underscores (\textbackslash{}\_) should not be used in URI & NoUnserscores & URI Design \\
    \bottomrule
    \end{tabularx}%
  \label{tab:restruler}%
\end{table}


Software quality refers not only to the absence of defects but also to architectural and structural design, which supports maintainability, usability, and long-term sustainability \cite{10741246}. In web applications, where systems interact through Web APIs, API design plays a crucial role: poor design can lead to misuse and increase maintenance costs \cite{di2023starting,10741246}.


The Representational State Transfer (REST) architectural style, introduced by Fielding \cite{fielding2000architectural}, has become the dominant paradigm for designing modern Web APIs. REST is not a protocol or a standard, but an architectural style defined by a set of constraints: statelessness, client-server separation, cacheability, a uniform interface, a layered system, and optional code-on-demand. APIs that fully adhere to these constraints are called REST APIs, and services that implement them are often referred to as RESTful \cite{masse2011rest}.

A REST API is composed of interlinked resources, and this assembly of resources forms the REST API resource model. Each resource is uniquely addressable via URIs and manipulated through a set of well-defined operations using standard HTTP methods (e.g., \texttt{GET}, \texttt{POST}, \texttt{PUT}, \texttt{DELETE}). In a RESTful system, the structure and semantics of URIs, HTTP methods, response codes, and payload formats should follow clear and consistent patterns \cite{pautasso2008restful}.

Designing a REST API, however, is not always straightforward. Although some guidelines are implied by the HTTP standard itself, others have emerged as best practices adopted by the community over time. These range from URI naming conventions to error handling and response formatting. Examples include placing verbs in URIs (e.g., \texttt{/createUser} instead of \texttt{/users}), misusing HTTP methods (e.g., using \texttt{GET} to delete a resource), or inconsistently applying status codes. As a result, REST API design is often viewed as a craft—balancing technical correctness, developer ergonomics, and consistency \cite{bogner2023restful}.

To reduce ambiguity and promote consistency in RESTful API design, numerous studies have attempted to formalize the abstract principles of REST into standardized design rules and best practices. These include both academic contributions, such as those by Petrillo et al. \cite{petrillo2016rest}, Palma et al. \cite{palma2022assessing}, and Massè et al. \cite{masse2011rest}. These works provide actionable guidance on resource naming, proper use of HTTP methods, status codes, and error handling, serving as foundational references for REST API design.

Building on foundational guidelines, several works have focused on automatically detecting REST API design rule violations that reduce usability and maintainability. Tools like \textsc{SARA} \cite{palma2017semantic} and \textsc{SOFA} \cite{moha2012specification} use rule-based and semantic analysis to uncover issues in URI structure, HTTP verb usage, and resource modeling. More recently, Bogner et al. introduced \textsc{RESTRuler} \cite{RestRuler}, a Java-based open-source tool that leverages static analysis to detect violations of 14 REST design rules, as shown in Table \ref{tab:restruler}. 
Other empirical studies have analyzed open-source APIs to assess real-world adherence to REST principles, revealing frequent deviations despite claims of RESTfulness \cite{bogner2023restful,kotstein2021restful}. 

Despite the importance of adhering to REST design principles, little attention has been paid to how these guidelines are understood and applied in educational settings. To date, no comprehensive studies have investigated the extent to which students are aware of these rules, nor how frequently they violate them when designing Web APIs. This lack of insight limits our ability to improve REST education and address misunderstandings at their root.






\section{Study Design}\label{sec:study}


The \textit{goal} of this study is to evaluate the quality of RESTful back-end components in full-stack web applications developed by students enrolled in a third-year Web Technologies course within a Computer Science Bachelor’s Degree program. As one of the final courses before graduation, it requires students to build complete web systems that integrate both front-end and back-end components.

This setting offers an opportunity to assess the technical competencies students acquire by the end of their academic journey. In particular, we focus on the design of REST APIs to identify recurring violations of established design principles. To this end, we pose the following research question:

\resquestion{1}{What are the most frequent REST API design rule violations in student-developed web applications?}

The results of this study are intended to help educators refine course content by highlighting specific design aspects that require greater emphasis. Furthermore, the findings offer insight into the level of API design maturity that companies can expect from recent graduates entering the job market.

\subsection{Web Technologies Course}

The projects analyzed in this study were developed as part of Web Technologies, a 6 ECTS third-year course in the B.Sc. in Computer Science at the University of Naples Federico II, taught by the last author of the paper. The course builds on prior knowledge of object-oriented programming, software engineering and computer networks, and covers the basis of modern web development. Involved students are expected to be familiar with object-oriented programming and core software engineering concepts such as the software development lifecycle, software architecture and design, and code quality. 

The course covers both client- and server-side web development, starting with foundational topics such as HTTP, modern HTML and CSS (including Flexbox, Grid, and responsive design), and advanced JavaScript (ES6+), focusing on DOM manipulation and asynchronous programming patterns. Server-side development progresses from traditional approaches like CGI and PHP to modern Node.js applications using Express, middleware, templating, and ORMs like Sequelize. REST APIs are a key focus, with dedicated lectures on authentication via JWT and design best practices \cite{masse2011rest}.

The course further explores modern front-end development with tooling, SPA architecture, TypeScript, and detailed coverage of the Angular framework, emphasizing components, dependency injection, and HTTP interceptors. Security and web testing are strongly emphasized, addressing vulnerabilities such as XSS, CSRF, and SQL injection, alongside testing strategies including end-to-end tests with tools like Playwright \cite{e2e-git-dataset-msr,10989035}.

Assessment combines a written exam and a practical project requiring students to design and develop an SPA with a RESTful backend supporting authentication and CRUD operations. The students were instructed that the project work would be graded primarily based on its functional correctness, but also taking into account overall software quality and adherence to established REST guidelines. Students may select technologies from the course or alternative frameworks, provided they respect SPA architectural principles.

\subsection{Collected Projects}
The web applications analyzed in this study were developed by students enrolled in the Spring 2024 edition of the Web Technologies course. From the initial submission of 52 projects, we conducted a manual validation process in which the first and third authors verified the satisfaction of dependencies, compilation, and execution capabilities. This quality control phase resulted in the exclusion of 12 projects, yielding a final dataset of 40 functionally complete and manually verified applications.

The technological composition of these projects demonstrates consistency in backend implementation; specifically, 39 projects (97.5\%) adopted JavaScript with the Express framework (consistent with the examples shown during the course), while one project (2.5\%) used Java with the Spring framework. 
Front-end implementations exhibited greater diversity, with Angular, which was also discussed in detail during the course, dominating as the framework of choice (33 projects, 82.5\%), followed by React (6 projects, 15\%), and a single instance using Vue (2.5\%). We have also observed notable differences between front-end and backend modules. The front-end modules demonstrate substantially larger codebases, averaging 2,099 lines of code (NCLOC) (Standard Deviation = 727) compared to 847 NCLOC (Standard Deviation = 603) for backend modules. A similar disparity can be observed for file organization, with front-end comprising an average of 66 files (Standard Deviation = 19), while the average for back-ends is 20 files (Standard Deviation = 11). 

\begin{figure}[!ht]
    \centering
    \includegraphics[width=0.7\linewidth]{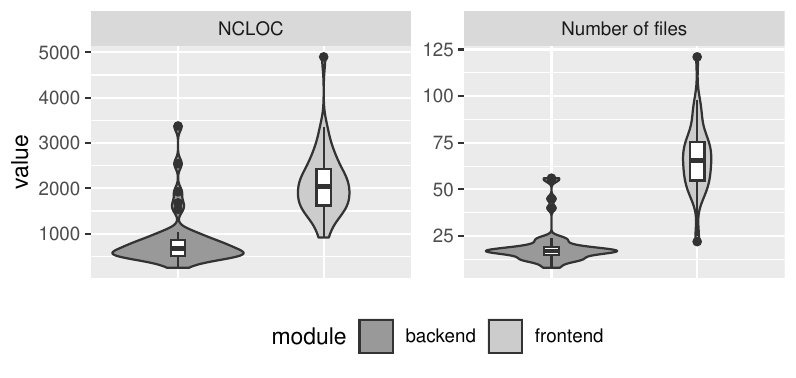}
    \caption{Project Statistics.}
    \label{fig:stats}
\end{figure}

The distribution of these metrics is reported in \Cref{fig:stats}. The Figure shows that while backend sizes cluster relatively tightly around the mean, front-end sizes exhibit greater variability, particularly in NCLOC. This pattern likely reflects both the inherent complexity of client-side development and the varying approaches students take in developing the front-end architecture.

\subsection{Automated Pipeline}
To evaluate the student submissions, we adopted an automated quality analysis pipeline inspired by the approach proposed by De Luca et al. \cite{de2024automatic}. The pipeline starts by extracting OpenAPI specifications from the back-end components of each project. For Express-based applications, we used \texttt{express-oas-generator} \cite{npmjsExpressoasgenerator}, while for Spring Boot projects, we employed \texttt{springdoc-openapi} \cite{springdocOpenAPILibrary}. These tools automatically inspect routing and controller logic to generate OpenAPI-compliant specifications. Once generated, the OpenAPI files were analyzed using \textsc{RESTRuler} \cite{RestRuler}, a tool that detects violations of widely accepted REST API design rules. We executed \textsc{RESTRuler} via its standalone JAR interface, producing a structured report for each project detailing the types and frequency of rule violations identified.

\section{Results for RQ$_1$: What are the most frequent REST API design rule violations in the student-developed web applications?}\label{sec:results}

The aggregated results on the detected REST design rule violations are reported in \Cref{tab:rule-violations}. 
\begin{table}
    \centering
    \renewcommand{\arraystretch}{1} 
    \setlength{\extrarowheight}{0pt}  
    \setlength{\aboverulesep}{0pt}
    \setlength{\belowrulesep}{0pt}
    \arrayrulecolor{black}
    \caption{Overview of rule violations in REST APIs, including occurrence count, affected projects, and percentage.}
    \label{tab:rule-violations}
    \begin{tabular}{|l|r|r|r|}
        \toprule
        \rowcolor{arsenic}
        \textcolor{white}{\textbf{Rule Identifier}} & \textcolor{white}{\textbf{Occurrences}} & \textcolor{white}{\textbf{\#Projects}} & \textcolor{white}{\textbf{(\%)}} \\
        \rowcolor{white}
        RC401 & 52 & 5 & 13 \\
        \rowcolor{gray!5}
        PluralNoun & 179 & 35 & 88 \\
        \rowcolor{white}
        SingularNoun & 48 & 12 & 30 \\
        \rowcolor{gray!5}
        NoTrailingSlash & 5 & 5 & 13 \\
        \rowcolor{white}
        VerbController & 1 & 1 & 3 \\
        \rowcolor{gray!5}
        NoCRUDNames & 61 & 11 & 28 \\
        \rowcolor{white}
        ContentType & 322 & 14 & 35 \\
        \rowcolor{gray!5}
         DescriptionType & 10 & 9 & 23 \\
        \rowcolor{white}
        ForwardSlash  & 3 & 3 & 8 \\
        \rowcolor{gray!5}
        NoTunnel & 28 & 4 & 10 \\
        \rowcolor{white}
        GETRetrieve & 201 & 33 & 83 \\
        \rowcolor{gray!5}
        Hyphens & 170 & 39 & 98 \\
        \rowcolor{white}
        Lowercase & 35 & 5 & 13 \\
        \rowcolor{gray!5}
        NoUnderscores& 5 & 3 & 8 \\
        \bottomrule
    \end{tabular}
\end{table}
These results show that notable patterns exist in how frequently certain best practices and design rules were overlooked across student projects. 

The most widespread issue was the \textit{missed use of hyphens to enhance readability} in endpoint paths, with the \textit{Hyphens} rule violated in 98\% of projects (39 out of 40). Similarly, pluralized nouns were not used for collection resource paths (\textit{PluralNoun}) in 88\% of projects (35 out of 40), while the \textit{GETRetrieve} rule, which ensured \texttt{GET} methods are used only for retrieval of resource representations, was violated in 83\% of projects (33 out of 40). These high percentages suggest that foundational REST conventions related to URI structure and HTTP method semantics were either not fully internalized or deliberately disregarded by the majority of students.

Other rules exhibited moderate prevalence. The \textit{ContentType} rule, which mandates explicit content-type headers, was violated in 35\% of projects (14 out of 40), indicating that a significant subset of students overlooked this aspect of API communication. Similarly, \textit{NoCRUDNames} (avoiding CRUD terms in URIs) and \textit{SingularNoun} (using singular nouns for document names) were violated in 28\% and 30\% of projects, respectively. These findings suggest that while some students adhered to these conventions, a substantial minority did not, potentially reflecting inconsistent emphasis during instruction or varying interpretations of REST guidelines.

Less common but still noteworthy were violations of rules like \textit{VerbController} (3\% of projects) and \textit{ForwardSlash} (8\% of projects). Their low prevalence implies that these were either more intuitive for students or less prone to ambiguity in teaching materials. However, even minor violations, such as \textit{NoTrailingSlash} and \textit{Lowercase} (each appearing in 13\% of projects), highlight areas where targeted instruction could further improve adherence.

Collectively, these results underscore that while students generally grasped high-level REST concepts, specific design principles, particularly those related to URI formatting and HTTP method usage, were frequently misapplied. The prevalence of violations in these areas suggests opportunities to refine pedagogical focus, perhaps through more explicit examples or automated linting tools during development.

\section{Threats to validity}\label{sec:threats}

\subsection{Construct Validity}

\textsc{RESTRuler}'s rule violations are based on a predefined set of REST design principles that may not encompass all real-world API design considerations. Additionally, since some projects required manual instrumentation to generate OpenAPI specifications for \textsc{RESTRuler}, inconsistencies in this process could introduce bias in the REST violation analysis.

\subsection{Internal Validity}
Manual steps involved in preparing projects for analysis—such as configuring or correcting OpenAPI generation—may have introduced variability in how \textsc{RESTRuler} was applied. Incomplete or improperly generated OpenAPI specifications might have caused some violations to be missed, affecting the completeness and comparability of results across projects.

\subsection{External Validity}

Our findings are based on a single cohort of students from a single university who followed a Web Technologies course. As such, the generalizability of the results to other educational contexts or professional settings may be limited. Additionally, the projects were developed under academic constraints (e.g., deadlines, grading criteria), which may not reflect real-world development practices.
\section{Conclusion and Future work}\label{sec:conclusions}

Our comprehensive analysis of 40 student-developed full-stack web applications revealed several patterns in REST API design violations that have important implications for computer science education. The findings demonstrate that while students generally master the functional aspects of web development, they struggle with quality-related concerns.
These findings suggest that architectural design rules deserve more explicit attention in undergraduate computer science curricula. Specifically, educators should consider integrating automated quality checks into course projects, providing targeted feedback on REST API design, and emphasizing the industrial relevance of these quality aspects.

For future works, we plan to extend our analysis to include software quality aspects such as code smells and maintainability issues by integrating tools like SonarQube into the evaluation process. Furthermore, we intend to investigate the effectiveness of various pedagogical interventions in improving students' software quality. This will involve comparing project quality before and after introducing dedicated lectures on software quality, incorporating continuous static analysis during development, and providing REST API design templates or checklists.


\bibliographystyle{plain}
\bibliography{bibliography}
\end{document}